\documentclass[aps,preprint]{revtex4}%
\usepackage{amsfonts}
\usepackage{amsmath}
\usepackage{amssymb}
\usepackage{subfigure}
\usepackage{graphicx}
\usepackage{array}%
\begin{document}
\preprint{CTP-SCU/2019004}
\title{Phase Structures and Transitions of Born-Infeld Black Holes in a Grand
Canonical Ensemble}
\author{Kangkai Liang}
\email{2016141221063@stu.scu.edu.cn}
\author{Peng Wang}
\email{pengw@scu.edu.cn}
\author{Houwen Wu}
\email{iverwu@scu.edu.cn}
\author{Mingtao Yang}
\email{2017141221040@stu.scu.edu.cn}
\affiliation{Center for Theoretical Physics, College of Physical Science and Technology,
Sichuan University, Chengdu, 610064, China}

\begin{abstract}
To make a Born-Infeld (BI) black hole thermally stable, we consider two types
of boundary conditions, i.e., the asymptotically anti-de Sitter (AdS) space
and a Dirichlet wall placed in the asymptotically flat space. The phase
structures and transitions of these two types of BI black holes, namely BI-AdS
black holes and BI black holes in a cavity, are investigated in a grand
canonical ensemble, where the temperature and the potential are fixed. For
BI-AdS black holes, the globally stable phases can be the thermal AdS space.
For small values of the potential, there is a Hawking-Page-like first order
phase transition between the BI-AdS black holes and the thermal-AdS space.
However, the phase transition becomes zeroth order when the values of the
potential are large enough. For BI black holes in a cavity, the globally
stable phases can be a naked singularity or an extremal black hole with the
horizon merging with the wall, which both are on the boundaries of the
physical parameter region. The thermal flat space is never globally preferred.
Besides a first order phase transition, there is a second order phase
transition between the globally stable phases. Thus, it shows that the phase
structures and transitions of BI black holes with these two different boundary
conditions have several dissimilarities.

\end{abstract}
\maketitle
\tableofcontents



\section{Introduction}

The study of black hole thermodynamics has continued to fascinate researchers
since the pioneering work
\cite{IN-Hawking:1974sw,IN-Bekenstein:1972tm,IN-Bekenstein:1973ur}, where
Hawking and Bekenstein found that black holes possess the temperature and the
entropy. However, it is well known that a Schwarzschild black hole in
asymptotically flat space is thermally unstable because of its negative
specific heat. To study black hole thermodynamics in a thermally stable
system, one can impose appropriate boundary conditions. For example, putting
black holes in the anti-de Sitter (AdS) space can make them thermally stable
since the AdS boundary acts as a reflecting wall for the Hawking radiation.
The investigations of the thermodynamic properties of AdS black holes have
come a long way since the discovery of the Hawking-Page phase transition
\cite{IN-Hawking:1982dh}, i.e., a phase transition between the thermal AdS
space and the Schwarzschild-AdS black hole. Later, with the advent of the
AdS/CFT correspondence
\cite{IN-Maldacena:1997re,IN-Gubser:1998bc,IN-Witten:1998qj}, there has been
much interest in studying the phase transitions of AdS black holes
\cite{IN-Witten:1998zw,IN-Chamblin:1999tk,IN-Chamblin:1999hg,IN-Caldarelli:1999xj,IN-Cai:2001dz,IN-Kubiznak:2012wp}%
. From the holographic perspective, we are eager to find out whether the
duality is independent of the details of the boundary conditions of the bulk
spacetime. It is therefore interesting to study the thermodynamics and phase
structures of black holes under different boundary conditions and look for
similarities or dissimilarities to the AdS case.

On the other hand, placing a Schwarzschild black hole in a cavity in the
asymptotically flat space, York showed that the black hole can be thermally
stable and has similar phase structure and transition to these of a
Schwarzschild-AdS black hole \cite{IN-York:1986it}. Specifically, the
Schwarzschild black hole in a cavity undergoes a Hawking-Page-like transition
to the thermal flat space as the temperature decreases. The thermodynamics and
phase structure of a Reissner-Nordstrom (RN) black hole in a cavity have been
studied in a grand canonical ensemble \cite{IN-Braden:1990hw} and a canonical
ensemble \cite{IN-Carlip:2003ne,IN-Lundgren:2006kt}, which showed that the
phase structures of the RN black hole in a cavity and the RN-AdS black hole
have extensive similarities. In a series of paper
\cite{IN-Lu:2010xt,IN-Wu:2011yu,IN-Lu:2012rm,IN-Lu:2013nt,IN-Zhou:2015yxa,IN-Xiao:2015bha}%
, the phase structures of various black brane systems in a cavity were
investigated in a grand canonical ensemble and a canonical ensemble, and it
was found that Hawking-Page-like or van der Waals-like phase transitions
always occur except for some special cases. In
\cite{IN-Basu:2016srp,IN-Peng:2017gss,IN-Peng:2017squ,IN-Peng:2018abh}, boson
stars and hairy black holes in a cavity were considered, and it showed that
the phase structure of the gravity system in a cavity is strikingly similar to
that of holographic superconductors in the AdS gravity. The stabilities of
solitons, stars and black holes in a cavity were also studied in
\cite{IN-Sanchis-Gual:2015lje,IN-Dolan:2015dha,IN-Ponglertsakul:2016wae,IN-Sanchis-Gual:2016tcm,IN-Ponglertsakul:2016anb,IN-Sanchis-Gual:2016ros,IN-Dias:2018zjg,IN-Dias:2018yey}%
, which showed that the nonlinear dynamical evolution of a charged black hole
in a cavity could end in a quasi-local hairy black hole. The thermodynamic
behavior of de Sitter black holes in a cavity has been discussed in the
extended phase space \cite{IN-Simovic:2018tdy}. Recently, McGough, Mezei and
Verlinde \cite{IN-McGough:2016lol} proposed that the holographic dual of
$T\bar{T}$ deformed CFT$_{\text{2}}$ is a finite region of AdS$_{\text{3}}$
with the wall at finite radial distance, which further motivates us to explore
the properties of a black hole in a cavity.

The Born-Infeld (BI) electrodynamics is a particular example of a nonlinear
electrodynamics, which is an effective model incorporating quantum corrections
to Maxwell electromagnetic theory. BI electrodynamics was first proposed to
smooth divergences of the electrostatic self-energy of point charges by
introducing a cutoff on electric fields \cite{IN-Born:1934gh}. Later, it is
realized that BI electrodynamics can emerge from the low energy limit of
string theory, which encodes the low-energy dynamics of D-branes. Coupling the
BI electrodynamics field to gravity, the BI black hole solution was first
obtained in \cite{IN-Dey:2004yt,IN-Cai:2004eh}. For the BI black holes in
asymptotically AdS space, the thermodynamic behavior and phase transitions
have been investigated in
\cite{IN-Fernando:2003tz,IN-Fernando:2006gh,IN-Banerjee:2010da,IN-Banerjee:2011cz,IN-Lala:2011np,IN-Banerjee:2012zm,IN-Gunasekaran:2012dq,IN-Zou:2013owa,IN-Azreg-Ainou:2014twa,IN-Hendi:2015hoa,IN-Zangeneh:2016fhy,IN-Zeng:2016sei,IN-Li:2016nll,IN-Hendi:2017oka,IN-Dehyadegari:2017hvd,IN-Tao:2017fsy,IN-Guo:2017bru,IN-Wang:2018xdz,IN-Momennia:2017hsc,
IN-Majhi:2016txt,IN-Bhattacharya:2017hfj,IN-Bhattacharya:2017nru}.
Specifically, the phase structures and transitions of 4D BI-AdS black holes in
a canonical ensemble were studied in
\cite{IN-Gunasekaran:2012dq,IN-Dehyadegari:2017hvd,IN-Wang:2018xdz}, which
showed that a reentrant phase transition was always observed in a certain
region of the parameter space. Meanwhile, the thermodynamics and phase
transitions in a grand canonical ensemble have been analyzed in
\cite{IN-Fernando:2006gh}, which showed that the system undergoes the first
and zeroth order phase transitions between the black hole solutions and the
thermal AdS space. On the other hand, by placing a BI black hole in a
spherical thermal cavity, we recently discussed the phase structures and
transitions of the canonical ensemble of this system \cite{IN-Wang:2019kxp},
which were found to have dissimilarities from these of the BI-AdS black holes.

In this paper, we study the phase structures and transitions of the grand
canonical ensemble of BI black holes using both asymptotically AdS and the
Dirichlet wall boundary conditions. So the gauge potential is fixed rather
than the charge on the boundaries in this paper. In the framework of the
AdS/CFT duality, the grand canonical ensemble is more relevant than the
canonical ensemble. Although the phase structures and transitions of BI-AdS
black holes in the grand ensemble have already been investigated in
\cite{IN-Fernando:2006gh}, we carry out the analysis in a more through way
with a broader survey of the parameter space. The phase diagrams in the
parameter space are obtained, which can be used to make a comparison with
these of BI black holes in a cavity. In the second part of this paper, we
analyze the phase structures and transitions of BI black hole in a cavity in
the grand canonical ensemble. We find that the thermal flat space, which is
the counterpart of the thermal AdS space in the BI-AdS case, can never be the
globally stable phase. Moreover, the system has no zeroth order transition,
but instead a second order transition occurs. It turns out that the results of
the BI black holes in a cavity and BI-AdS black holes have several dissimilarities.

The rest of this paper is organized as follows. In section \ref{Sec:BIAdSBH},
we study the phase structures and transitions of BI-AdS black holes and give
the phase diagrams, e.g., FIGs. \ref{fig:RP} and \ref{fig:PD}. In section
\ref{Sec:BIBHC}, we discuss the phase structures and transitions of BI black
holes in a cavity. The related phase diagrams are given in FIGs. \ref{fig:CRP}
and \ref{fig:CPD}, from which one can read the phase structures and
transitions. Section \ref{Sec:Con} is devoted to our discussion and conclusion.

\section{Born-Infeld AdS Black Holes}

\label{Sec:BIAdSBH}

In this section, we consider the phase structures and transitions of BI-AdS
black holes in a grand canonical ensemble. The action of a $\left(
3+1\right)  $ dimensional model of gravity coupled to a Born-Infeld
electromagnetic field $A_{\mu}$ is%
\begin{equation}
\mathcal{S}=\int d^{4}x\sqrt{-g}\left[  R-2\Lambda+\mathcal{L}_{\text{BI}%
}\left(  F\right)  \right]  , \label{eq:BIAdSAction}%
\end{equation}
where the cosmological constant $\Lambda=-3/l^{2}$, and we take $16\pi G=1$
for simplicity. The Born-Infeld electrodynamics Lagrangian density is%
\[
\mathcal{L}_{\text{BI}}\left(  F\right)  =-\frac{1}{a}\left(  1-\sqrt
{1-aF^{\mu\nu}F_{\mu\nu}/2}\right)  ,
\]
where $F_{\mu\nu}=\partial_{\mu}A_{\nu}-\partial_{\nu}A_{\mu}$, and the
Born-Infeld parameter $a$ is related to the string tension $\alpha^{\prime}$
as $a=\left(  2\pi\alpha^{\prime}\right)  ^{2}>0$. When $a\rightarrow0$,
$\mathcal{L}_{\text{BI}}\left(  F\right)  $ reduces to the Lagrangian of the
Maxwell field. The Born-Infeld AdS black hole solution was obtained in
\cite{IN-Dey:2004yt,IN-Cai:2004eh}:%
\begin{align}
ds^{2}  &  =-f\left(  r\right)  dt^{2}+\frac{dr^{2}}{f\left(  r\right)
}+r^{2}\left(  d\theta^{2}+\sin^{2}\theta d\phi^{2}\right)  \text{,}%
\nonumber\\
A  &  =A_{t}\left(  r\right)  dt\text{.}%
\end{align}
where%
\begin{align}
f\left(  r\right)   &  =1-\frac{M}{8\pi r}+\frac{r^{2}}{l^{2}}-\frac{Q^{2}%
}{6\sqrt{r^{4}+aQ^{2}}+6r^{2}}+\frac{Q^{2}}{3r^{2}}\text{ }_{2}F_{1}\left(
\frac{1}{4},\frac{1}{2},\frac{5}{4};-\frac{aQ^{2}}{r^{4}}\right)
\text{,}\nonumber\\
A_{t}^{\prime}\left(  r\right)   &  =\frac{Q}{\sqrt{r^{4}+aQ^{2}}}.
\end{align}
Here $M$ and $Q$ are the mass and the charge of the back hole, respectively,
and $_{2}F_{1}\left(  a,b,c;x\right)  $ is the hypergeometric function.

At the horizon $r=r_{+}$, one has that $f\left(  r_{+}\right)  =0$, and the
Hawking temperature is given by%
\begin{equation}
T=\frac{f^{\prime}\left(  r_{+}\right)  }{4\pi}=\frac{1}{4\pi r_{+}}\left(
1-\frac{1}{2}\frac{Q^{2}}{r_{+}^{2}+\sqrt{r_{+}^{4}+aQ^{2}}}\right)  \text{.}
\label{eq:BIAdST}%
\end{equation}
Requiring $A_{t}\left(  r\right)  $ at the horizon to be zero, it can show
that the gauge potential measured with respect to the horizon is%
\begin{equation}
\Phi=4\pi A_{t}\left(  \infty\right)  =\frac{4\pi Q}{r_{+}}\text{ }_{2}%
F_{1}\left(  \frac{1}{4},\frac{1}{2},\frac{5}{4};-\frac{aQ^{2}}{r_{+}^{4}%
}\right)  . \label{eq:BIAdSPotential}%
\end{equation}
In the limit of $r_{+}\rightarrow+\infty$, BI-AdS black holes would reduce to
RN-AdS black holes, and we find that%
\begin{equation}
\Phi\left(  r_{+},Q,a\right)  \sim\frac{4\pi Q}{r_{+}}\text{ and }T\sim
\frac{3r_{+}}{4\pi l^{2}}\text{.}%
\end{equation}
As $r_{+}\rightarrow0$, eqns. $\left(  \ref{eq:BIAdST}\right)  $ and $\left(
\ref{eq:BIAdSPotential}\right)  $ gives
\begin{equation}
\Phi\sim\sqrt{\frac{Q}{2\sqrt{a}}}\Phi_{c}\text{ and }T\sim\frac{1}{4\pi
r_{+}}\left(  1-\frac{\Phi^{2}}{\Phi_{c}^{2}}\right)  ,
\end{equation}
where $\Phi_{c}\equiv4\sqrt{2\pi}\Gamma\left(  \frac{1}{4}\right)
\Gamma\left(  \frac{5}{4}\right)  \sim32.95$. So when $\Phi>\Phi_{c}$,
$T\rightarrow-\infty$ as $r_{+}\rightarrow0$, which means that $r_{+}$ has a
nonzero minimum value. On the other hand, $T\rightarrow+\infty$ as
$r_{+}\rightarrow0$ for $\Phi<\Phi_{c}$, and hence $r_{+}$ can go to zero in
this case.

To study the phase structures and transitions, we need to consider the free
energy of the black hole. The free energy of a BI-AdS black hole in a
canonical ensemble was obtained by computing the Euclidean action in
\cite{IN-Wang:2018xdz}, where an extra boundary term $\mathcal{S}%
_{\text{surf}}$ was introduced to keep the charge fixed instead of the
potential. However for the grand canonical ensemble, $\mathcal{S}%
_{\text{surf}}$ is not needed any more. Excluding the contribution of
$\mathcal{S}_{\text{surf}}$, the computation of the Euclidean action in
\cite{IN-Wang:2018xdz} then gives the free energy of the BI-AdS black hole in
the grand canonical ensemble:%
\begin{equation}
F=M-TS-Q\Phi,
\end{equation}
where $S=16\pi^{2}r_{+}^{2}$ is the entropy of the black hole. For the later
convenience, we can express quantities in units of $l$:%
\begin{equation}
\tilde{T}=Tl\text{, }\tilde{r}_{+}=r_{+}/l\text{, }\tilde{Q}=Q/l\text{,
}\tilde{a}=al^{-2}\text{ and }\tilde{F}\equiv F/l\text{.}%
\end{equation}
Note that the potential $\Phi$ is dimensionless.

\begin{figure}[tb]
\begin{center}
\includegraphics[width=0.43\textwidth]{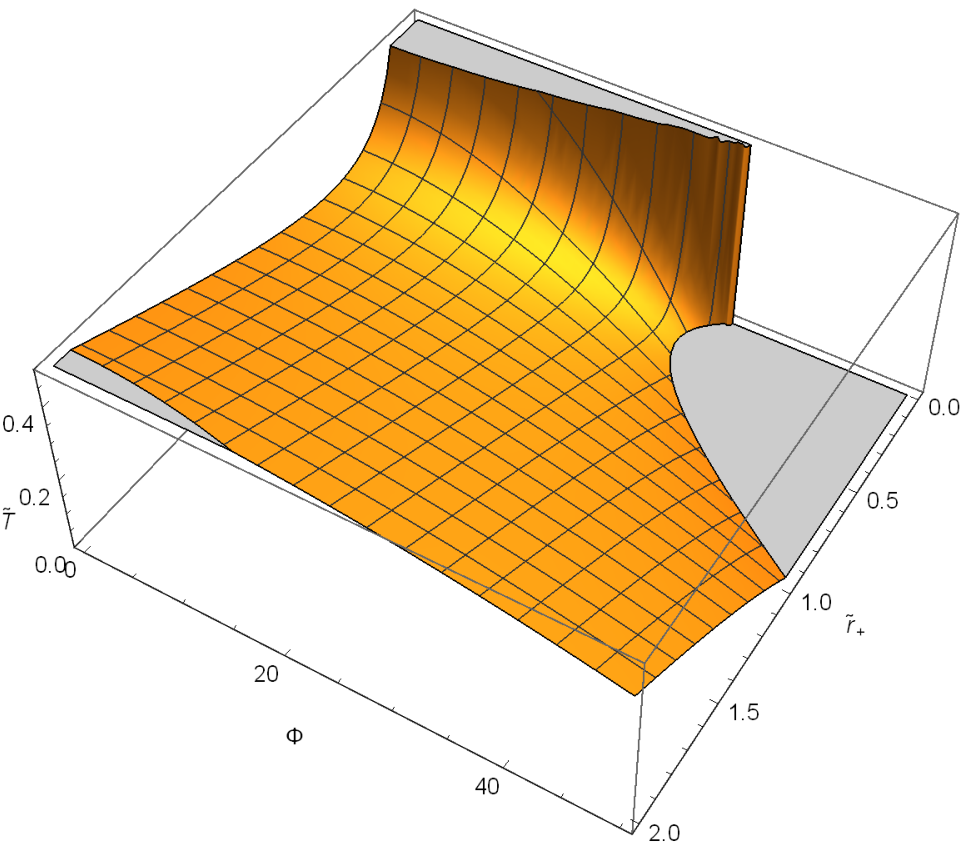}
\includegraphics[width=0.52\textwidth]{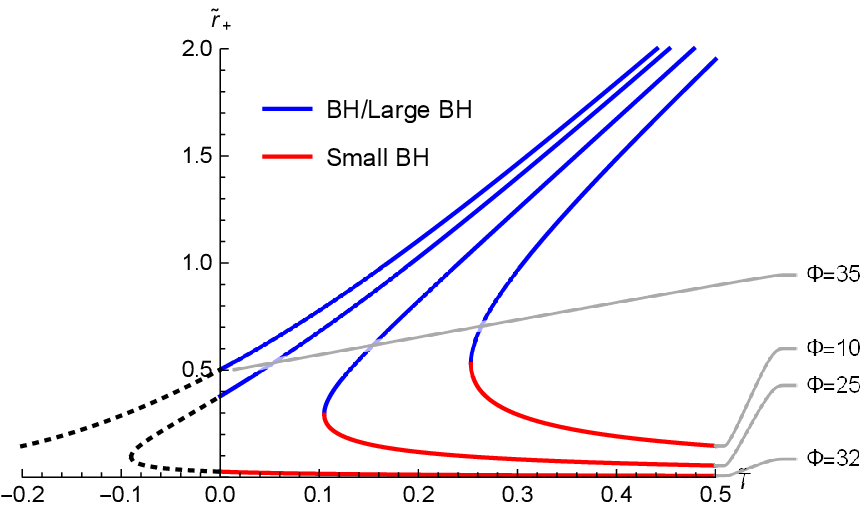}
\end{center}
\caption{{\footnotesize Plots of $\tilde{T}(\tilde{r}_{+},\Phi)$ and
$\tilde{r}_{+}(\tilde{T},\Phi)$ for BI-AdS black holes with $\tilde{a}=0.1$.
\textbf{Left Panel}: Plot of $\tilde{T}$ as a function of $\tilde{r}_{+}$ and
$\Phi$. $\tilde{T}$ is negative in the gray area. \textbf{Right Panel}: Plot
of $\tilde{r}_{+}$ against $\tilde{T}$ for various values of $\Phi$. Since
thermally stable phases have $\partial\tilde{r}_{+}/\partial\tilde{T}>0$, the
BI-AdS black holes on blue/red branches of the $\tilde{r}_{+}(\tilde{T},\Phi)$
curves are thermally stable/unstable. We denote the blue branches by Large BH
(or BH if there is only one branch) and red branches by Small BH. The black
holes on the black dashed branches have negative temperature, which are
unphysical. }}%
\label{fig:PlotT}%
\end{figure}

To find the phase structures of the black hole, one needs to use eqns.
$\left(  \ref{eq:BIAdST}\right)  $ and $\left(  \ref{eq:BIAdSPotential}%
\right)  $ to express the horizon radius $\tilde{r}_{+}$ in terms of the
temperature $\tilde{T}$ and the potential $\Phi$: $\tilde{r}_{+}=\tilde{r}%
_{+}(\tilde{T},\Phi)$. When $\Phi<\Phi_{c}$, $T\rightarrow+\infty$ in the
limits of $r_{+}\rightarrow0$ and $r_{+}\rightarrow+\infty$, which implies
that $\tilde{r}_{+}(\tilde{T},\Phi)$ are multivalued. In the left panel of
FIG. \ref{fig:PlotT}, we plot $\tilde{T}$ as a function of $\tilde{r}_{+}$ and
$\Phi$ with $\tilde{a}=0.1$. We plot $\tilde{r}_{+}$ against $\tilde{T}$ for
various values of $\Phi$ with $\tilde{a}=0.1$ in the right panel of FIG.
\ref{fig:PlotT}. When $\Phi=10$, $25$ and $32$, there are two family of black
holes of different sizes with the same values of $\tilde{T}$ and $\Phi$: Small
BH and and Large BH. When $\Phi=35>\Phi_{c}$, there is only one branch: BH. To
consider the thermodynamic stabilities against thermal fluctuations, we
consider the specific heat at constant potential:%
\begin{equation}
C_{\Phi}=T\left(  \frac{\partial S}{\partial T}\right)  _{\Phi}=32l^{2}\pi
^{2}\tilde{r}_{+}\tilde{T}\frac{\partial\tilde{r}_{+}(\tilde{T},\Phi
)}{\partial\tilde{T}}.
\end{equation}
The thermal stable black holes have $C_{\Phi}\geq0$, which means
$\partial\tilde{r}_{+}/\partial\tilde{T}>0$. So the BH/Large BH branches in
FIG. \ref{fig:PlotT} are thermally stable. To discuss the phase transitions of
the black hole, we need to calculate the free energies of different branches
and compare them. Moreover, the thermal AdS space with a constant gauge
potential is also a classical solution of the action $\left(
\ref{eq:BIAdSAction}\right)  $. Therefore, the thermal AdS space is also
considered for the phase transitions in the grand canonical ensemble.

\begin{figure}[tb]
\begin{center}
\includegraphics[width=0.5\textwidth]{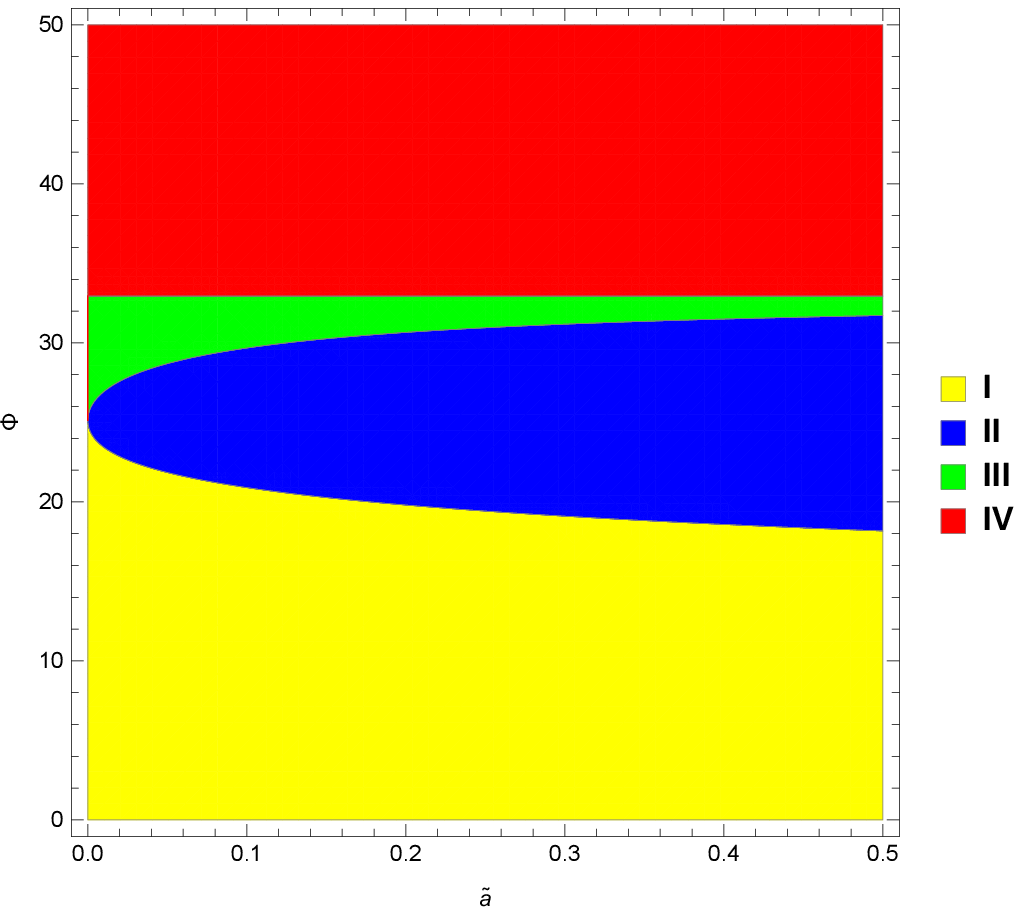}
\end{center}
\caption{{\footnotesize The four regions in the $\tilde{a}$-$\Phi$ phase space
of BI-AdS black holes, each of which possesses distinct behavior of the phase
structures and transitions. Varying the temperature, a first order LBH/Thermal
AdS phase transition occurs in Regions I while a zeroth order LBH/Thermal AdS
phase transition occurs in Regions II. There are no phase transitions in
Regions III and IV. }}%
\label{fig:RP}%
\end{figure}

\begin{figure}[ptb]
\begin{center}
\subfigure[{~\scriptsize Region I: $\tilde{a}=0.1$ and $\Phi=10$. As $\tilde{T}$ increases from
zero, there is a first order phase transition occurring at the black dot from the thermal AdS space to Large BH.}]{
\includegraphics[width=0.45\textwidth]{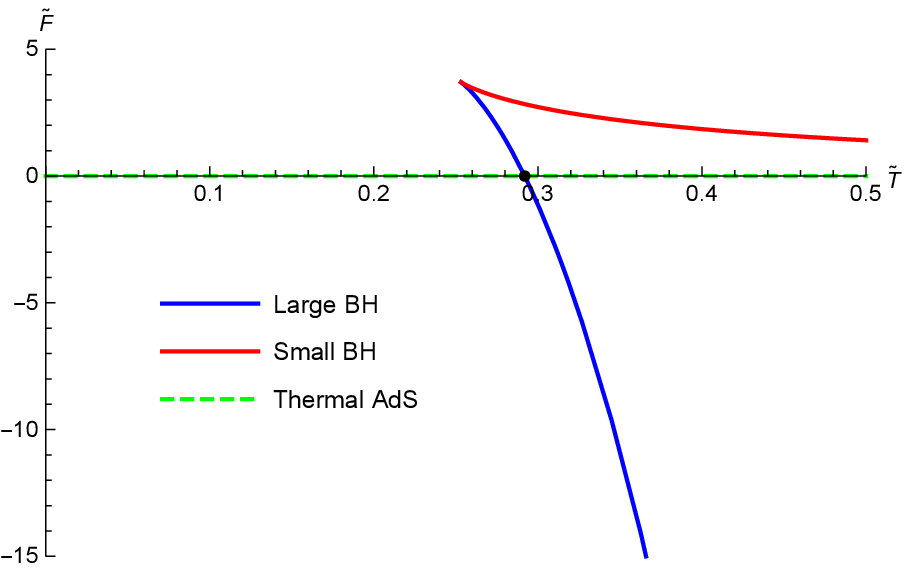}\label{fig:R:a}}
\subfigure[{~ \scriptsize Region II: $\tilde{a}=0.1$ and $\Phi=25$. As $\tilde{T}$ increases from
zero, there is a zeroth order phase transition occurring along the black dotted line from the thermal AdS space to Large BH.}]{
\includegraphics[width=0.45\textwidth]{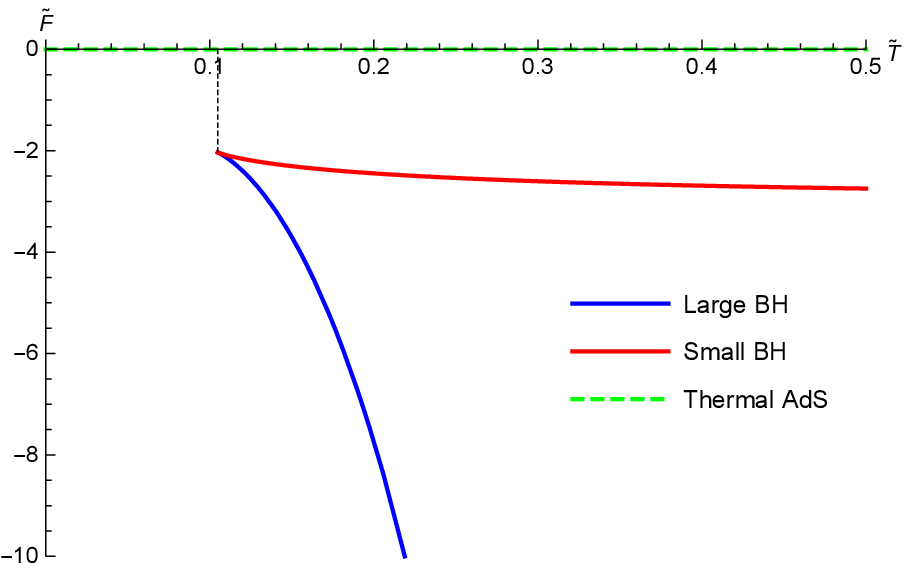}\label{fig:R:b}}
\subfigure[{~ \scriptsize Region III: $\tilde{a}=0.1$ and $\Phi=32$. There is no phase transition.}]{
\includegraphics[width=0.45\textwidth]{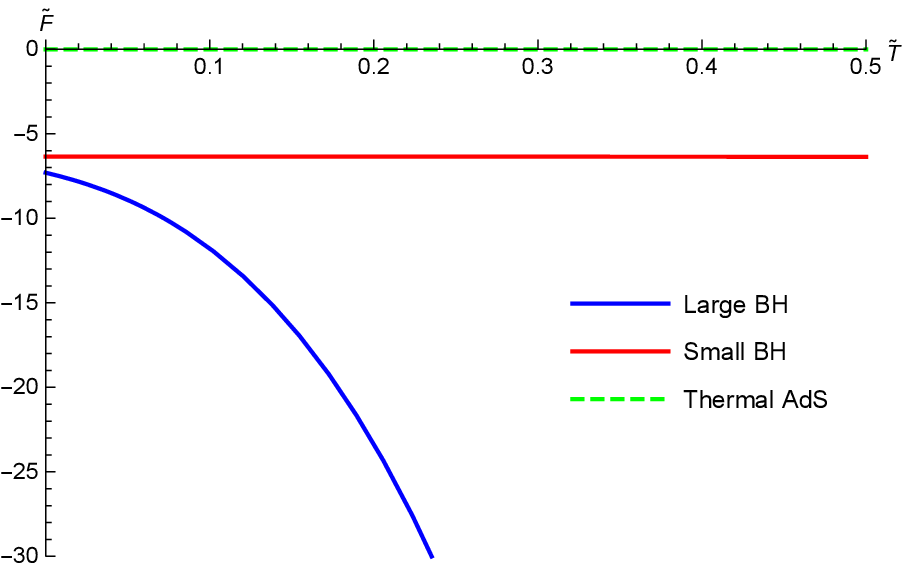}\label{fig:R:c}}
\subfigure[{~ \scriptsize Region IV: $\tilde{a}=0.1$ and $\Phi=35$. There is no phase transition.}]{
\includegraphics[width=0.45\textwidth]{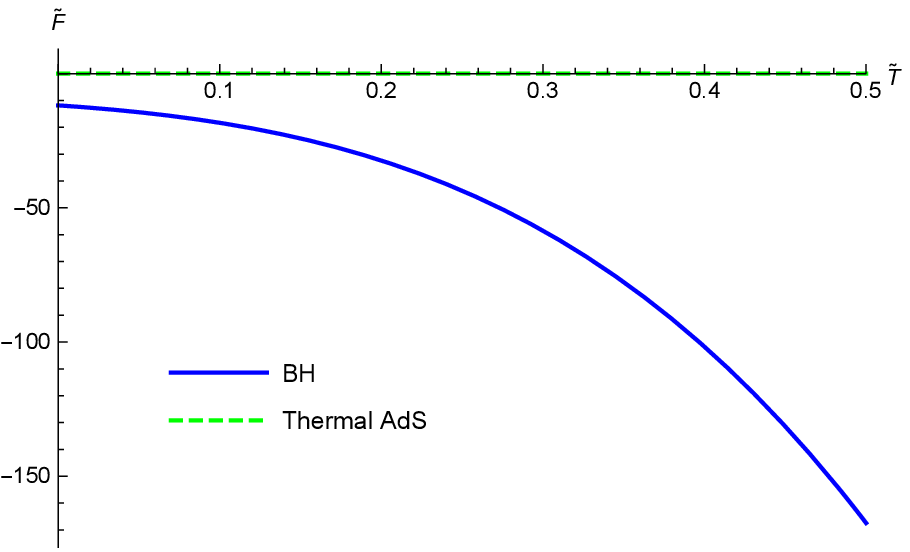}\label{fig:R:d}}
\end{center}
\caption{{\footnotesize Plots of the free energy $\tilde{F}$ against the
temperature $\tilde{T}$ for BI-AdS black holes in Regions I, II, III and IV.
The black holes on the blue branches are thermally stable.}}%
\end{figure}

We find that there are four regions in the $\tilde{a}$-$\Phi$ phase space of
the BI-AdS black holes, in each of which the black holes have different
behavior of the branches of $\tilde{r}_{+}(\tilde{T},\Phi)$ and phase
structure. These four regions of the $\tilde{a}$-$\Phi$ phase space are mapped
in FIG. \ref{fig:RP}. In what follows, we discuss the phase structures and
transitions in the four regions:

\begin{itemize}
\item Region I: The temperature of a BI-AdS black hole in this region has a
positive minimum value $\tilde{T}_{\min}$. For $\tilde{T}\geq\tilde{T}_{\min}%
$, there are two branches of black holes: Small BH and Large BH.\ The free
energies of the two branches with $\tilde{a}=0.1$ and $\Phi=10$ and the
thermal AdS space are plotted in FIG. \ref{fig:R:a}. The Large BH branch
always has lower free energy than the Small BH branch. The thermal AdS space
is the only phase when $\tilde{T}<\tilde{T}_{\min}$. At $\tilde{T}_{\min}$,
the black hole appears, and its free energy is larger than that of the thermal
AdS space. As $\tilde{T}$ increases from $\tilde{T}_{\min}$, the free energy
of Large BH decreases while that of the thermal AdS space is constant. They
cross each other at some point, where a first-order transition occurs, and
Large BH then becomes globally stable.

\item Region II: As in Region I, only the thermal AdS space exists for
$\tilde{T}<\tilde{T}_{\min}$, and\ the BI-AdS black hole appears and has two
branches for $\tilde{T}>\tilde{T}_{\min}$. The free energies of the two
branches with $\tilde{a}=0.1$ and $\Phi=25$ and the thermal AdS space are
plotted in FIG. \ref{fig:R:b}. However, at $\tilde{T}=\tilde{T}_{\min}$, the
free energy of the black hole is smaller than that of the thermal AdS space.
So there is a finite jump in the free energy at $\tilde{T}=\tilde{T}_{\min}$
leading to a zeroth order phase transition from the thermal AdS space to Large BH.

\item Region III: In this region, the BI-AdS black holes can exist for all
non-negative values of $\tilde{T}$, which have Large BH and Small BH branches.
The free energies of the two branches with $\tilde{a}=0.1$ and $\Phi=32$ and
the thermal AdS space are plotted in FIG. \ref{fig:R:c}. It shows that Large
BH always has the smallest free energy. So there is no phase transition, and
the global stable phase is Large BH.

\item Region IV: Since $\Phi>\Phi_{c}$ in this region, there is only one
branch for the BI-AdS black holes. The free energies of the black hole with
$\tilde{a}=0.1$ and $\Phi=35$ and the thermal AdS space are plotted in FIG.
\ref{fig:R:d}. As in Region III, Large BH is the only global stable phase, and
there is no phase transition.
\end{itemize}

\begin{figure}[tb]
\begin{center}
\includegraphics[width=0.48\textwidth]{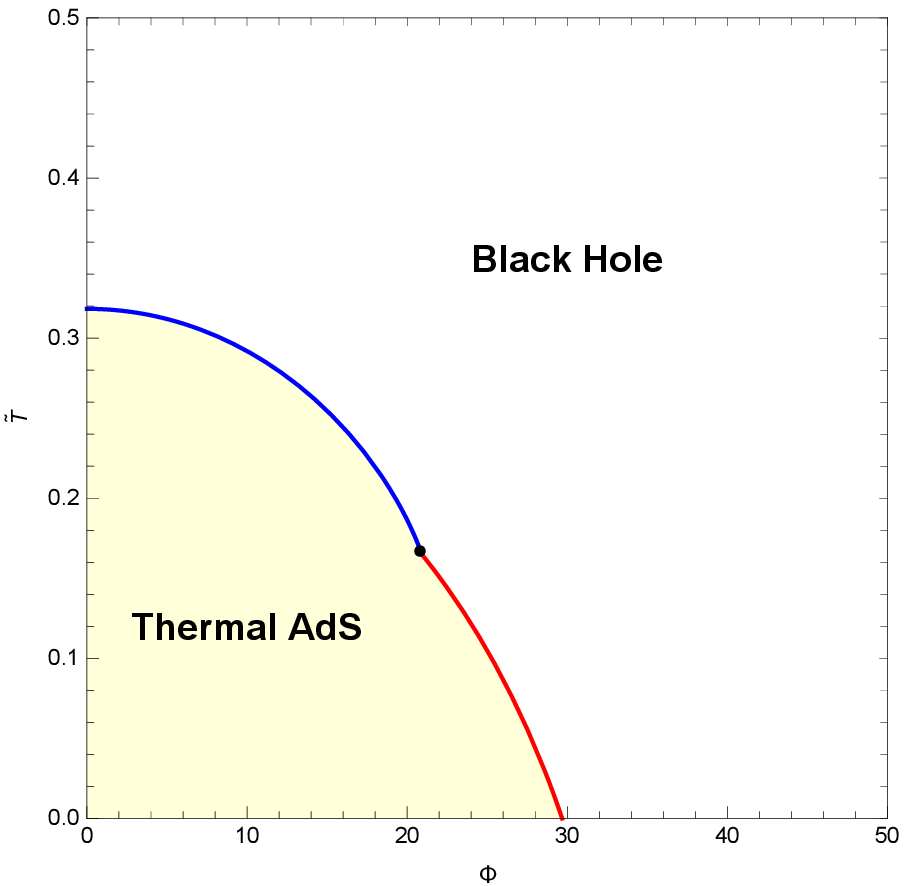}
\includegraphics[width=0.48\textwidth]{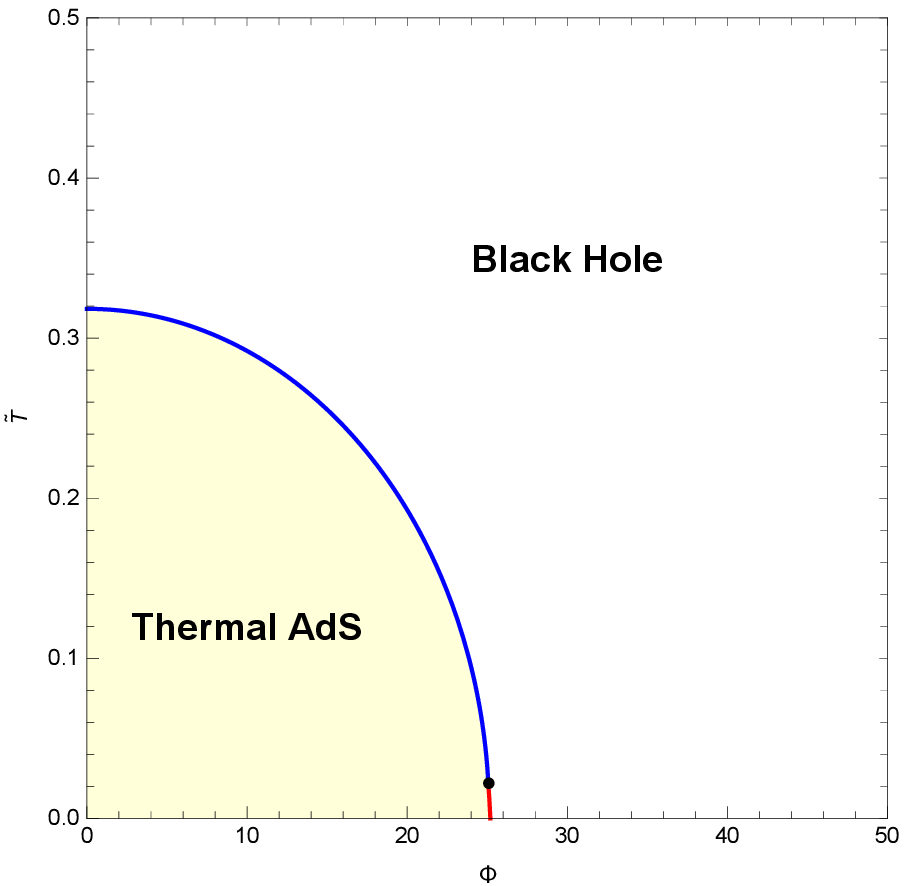}
\end{center}
\caption{{\footnotesize The phase diagrams of BI-AdS black holes in the $\Phi
$-$\tilde{T}$ phase space. \textbf{Left Panel}: $\tilde{a}=0.1$. \textbf{Right
Panel}: $\tilde{a}=10^{-5}$. The first/zeroth order phase transition lines
separating the black holes and the thermal AdS space are displayed by the
blue/red lines. The first and zeroth order phase transition lines meet and
terminate at the black dots.} }%
\label{fig:PD}%
\end{figure}

The phase diagram in the $\Phi$-$\tilde{T}$ space of the BI-AdS black hole
with $\tilde{a}=0.1$ is displayed in the left panel of FIG. \ref{fig:PD}.
There is a BH/Thermal AdS first order phase transition line for some range of
$\Phi$ and a BH/Thermal AdS zeroth order phase transition line for larger
values of $\Phi$. These two phase transition lines meet and terminate at the
black dot. Here, we simply use BH to denote Large BH without causing any
confusion. The phase diagram of the BI-AdS black hole with $\tilde{a}=10^{-5}$
is displayed in the right panel of FIG. \ref{fig:PD}, which is similar to the
$\tilde{a}=0.1$ case. It is noteworthy that the zeroth order phase transition
line becomes shorter for a smaller value of $\tilde{a}$. For a RN-AdS black
hole, which has $\tilde{a}=0$, there is no zeroth order phase transition
\cite{IN-Fernando:2006gh}.

\section{Born-Infeld Black Holes in a Cavity}

\label{Sec:BIBHC}

In this section, we consider a thermodynamic system with Born-Infeld
electrodynamics charged black holes inside a cavity, on the boundary of which
the temperature and the potential are fixed. On a $\left(  3+1\right)  $
dimensional spacetime manifold $\mathcal{M}$ with a time-like boundary
$\partial\mathcal{M}$, the action is given by%
\begin{equation}
\mathcal{S}=\int_{\mathcal{M}}d^{4}x\sqrt{-g}\left[  R+\mathcal{L}_{\text{BI}%
}\left(  F\right)  \right]  -2\int_{\partial\mathcal{M}}d^{3}x\sqrt{-\gamma
}\left(  K-K_{0}\right)  , \label{eq:CavityAction}%
\end{equation}
where $K$ is the extrinsic curvature, $\gamma$ is the metric on the boundary,
and $K_{0}$ is a subtraction term to make the boundary term vanish in flat
spacetime. The BI black hole solution of the action $\left(
\ref{eq:CavityAction}\right)  $ is \cite{IN-Wang:2019kxp}%
\begin{align}
ds^{2}  &  =-f\left(  r\right)  dt^{2}+\frac{dr^{2}}{f\left(  r\right)
}+r^{2}\left(  d\theta^{2}+\sin^{2}\theta d\phi^{2}\right)  \text{,}%
\nonumber\\
A  &  =A_{t}\left(  r\right)  dt\text{,} \label{eq:ansatz}%
\end{align}
where%
\begin{align}
f\left(  r\right)   &  =1-\frac{M}{8\pi r}-\frac{Q^{2}}{6\sqrt{r^{4}+aQ^{2}%
}+6r^{2}}+\frac{Q^{2}}{3r^{2}}\text{ }_{2}F_{1}\left(  \frac{1}{4},\frac{1}%
{2},\frac{5}{4};-\frac{aQ^{2}}{r^{4}}\right)  \text{,}\nonumber\\
A_{t}^{\prime}\left(  r\right)   &  =\frac{Q}{\sqrt{r^{4}+aQ^{2}}}.
\end{align}
Here $M$ and $Q$ are the mass and the charge of the back hole, respectively.
Note that $M$ plays no role in our paper since we always use the horizon
radius $r_{+}$ to eliminate $M$.

Suppose that the wall of the cavity enclosing the BI black holes is at
$r=r_{B}$, and the wall is maintained at a temperature of $T$ and a gauge
potential of $\Phi$, where we assume that $\Phi>0$ without loss of generality.
For this system, the Euclidean continuation of the action $\mathcal{S}$ was
calculated in \cite{IN-Wang:2019kxp}:%
\begin{equation}
\mathcal{S}^{E}=\frac{16\pi r_{B}}{T}\left[  1-\sqrt{f\left(  r_{B}\right)
}\right]  -S-\frac{Q\Phi}{T},
\end{equation}
where $S=16\pi^{2}r_{+}^{2}$ is the entropy of the black hole. In the
semiclassical approximation, the free energy $F$ is related to $\mathcal{S}%
^{E}$ by%
\begin{equation}
F=T\mathcal{S}^{E}.
\end{equation}
Expressing the mass $M$ in terms of the horizon radius $r_{+}$, one finds that
the free energy $F$ is a function of the temperature $T$, the potential $\Phi
$, the charge $Q$, the cavity radius $r_{B}$ and the horizon radius $r_{+}$:%
\begin{equation}
F=F\left(  r_{+},Q;T,\Phi,r_{B}\right)  , \label{eq:F(rplus)}%
\end{equation}
where $T$, $\Phi$ and $r_{B}$ are parameters of the grand canonical ensemble.
The locally stationary points of the free energy $F$ can be determined by
extremizing $F\left(  r_{+},Q;T,\Phi,r_{B}\right)  $ with respect to $r_{+}$
and $Q$:%
\begin{align}
\frac{dF\left(  r_{+},Q;T,\Phi,r_{B}\right)  }{dr_{+}}  &  =0\Longrightarrow
T=\frac{T_{h}}{\sqrt{f\left(  r_{B}\right)  }}\text{,}\nonumber\\
\frac{dF\left(  r_{+},Q;T,\Phi,r_{B}\right)  }{dQ}  &  =0\Longrightarrow
\Phi=\frac{4\pi A_{t}\left(  r_{B}\right)  }{\sqrt{f\left(  r_{B}\right)  }%
}\text{,} \label{eq:LE}%
\end{align}
where
\begin{equation}
T_{h}=\frac{1}{4\pi r_{+}}\left(  1-\frac{1}{2}\frac{Q^{2}}{r_{+}^{2}%
+\sqrt{r_{+}^{4}+aQ^{2}}}\right)  \text{,} \label{eq:HT}%
\end{equation}
is the Hawking temperature of the black hole. Usually, it is convenient to
express quantities in units of $r_{B}$:%
\begin{equation}
x\equiv\frac{r_{+}}{r_{B}}\text{, }\tilde{Q}\equiv\frac{Q}{r_{B}}\text{,
}\tilde{a}\equiv\frac{a}{r_{B}^{2}}\text{, }\tilde{T}\equiv r_{B}T\text{ and
}\tilde{F}=\frac{F}{16\pi r_{B}}\text{.}%
\end{equation}
The potential $\Phi$ is dimensionless. In terms of $x$ and tilde quantities,
$f\left(  r_{B}\right)  $ and $\tilde{F}$ can be expressed as%
\begin{align}
f\left(  x\right)   &  =1-x+\frac{x\tilde{Q}^{2}}{6\sqrt{x^{4}+\tilde{a}%
\tilde{Q}^{2}}+6x^{2}}-\frac{\tilde{Q}^{2}}{3x}\text{ }_{2}F_{1}\left(
\frac{1}{4},\frac{1}{2},\frac{5}{4};-\frac{\tilde{a}\tilde{Q}^{2}}{x^{4}%
}\right) \nonumber\\
&  -\frac{\tilde{Q}^{2}}{6\sqrt{1+\tilde{a}\tilde{Q}^{2}}+6}+\frac{\tilde
{Q}^{2}}{3}\text{ }_{2}F_{1}\left(  \frac{1}{4},\frac{1}{2},\frac{5}%
{4};-\tilde{a}\tilde{Q}^{2}\right)  ,\\
\tilde{F}  &  =\tilde{F}\left(  x,\tilde{Q};\tilde{T},\Phi\right)
=1-\sqrt{f\left(  x\right)  }-\pi x^{2}\tilde{T}-\frac{\tilde{Q}\Phi}{16\pi
},\nonumber
\end{align}
respectively.

\begin{figure}[tb]
\begin{center}
\includegraphics[width=0.42\textwidth]{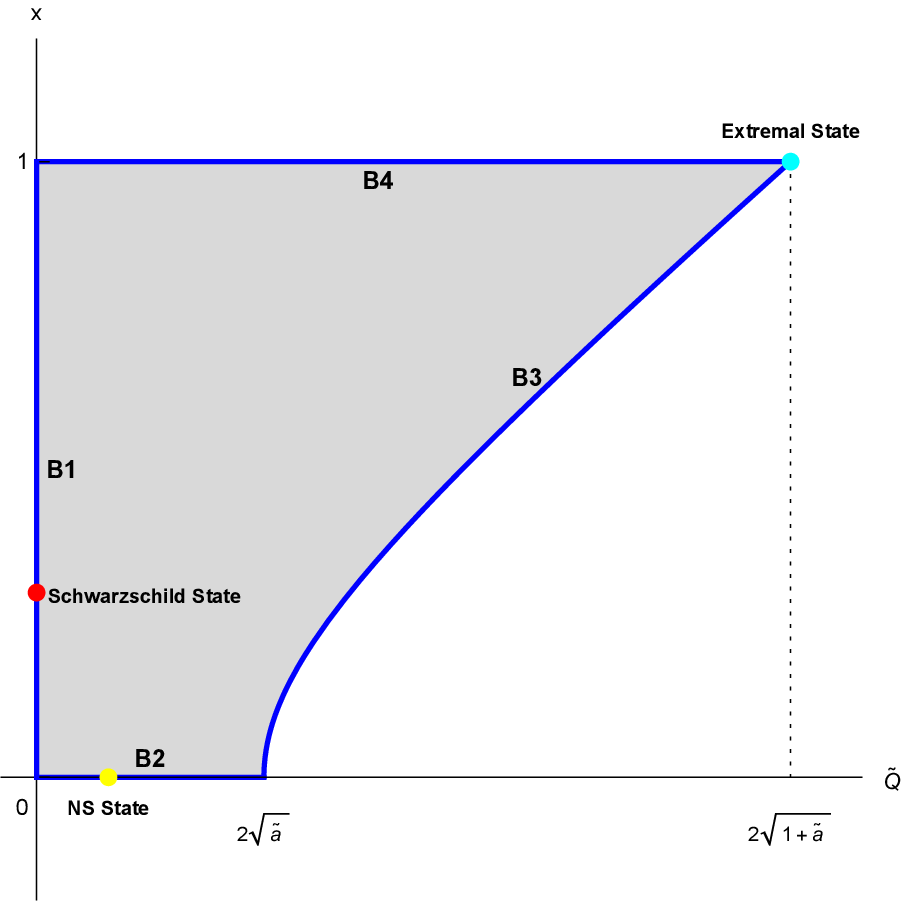}
\includegraphics[width=0.53\textwidth]{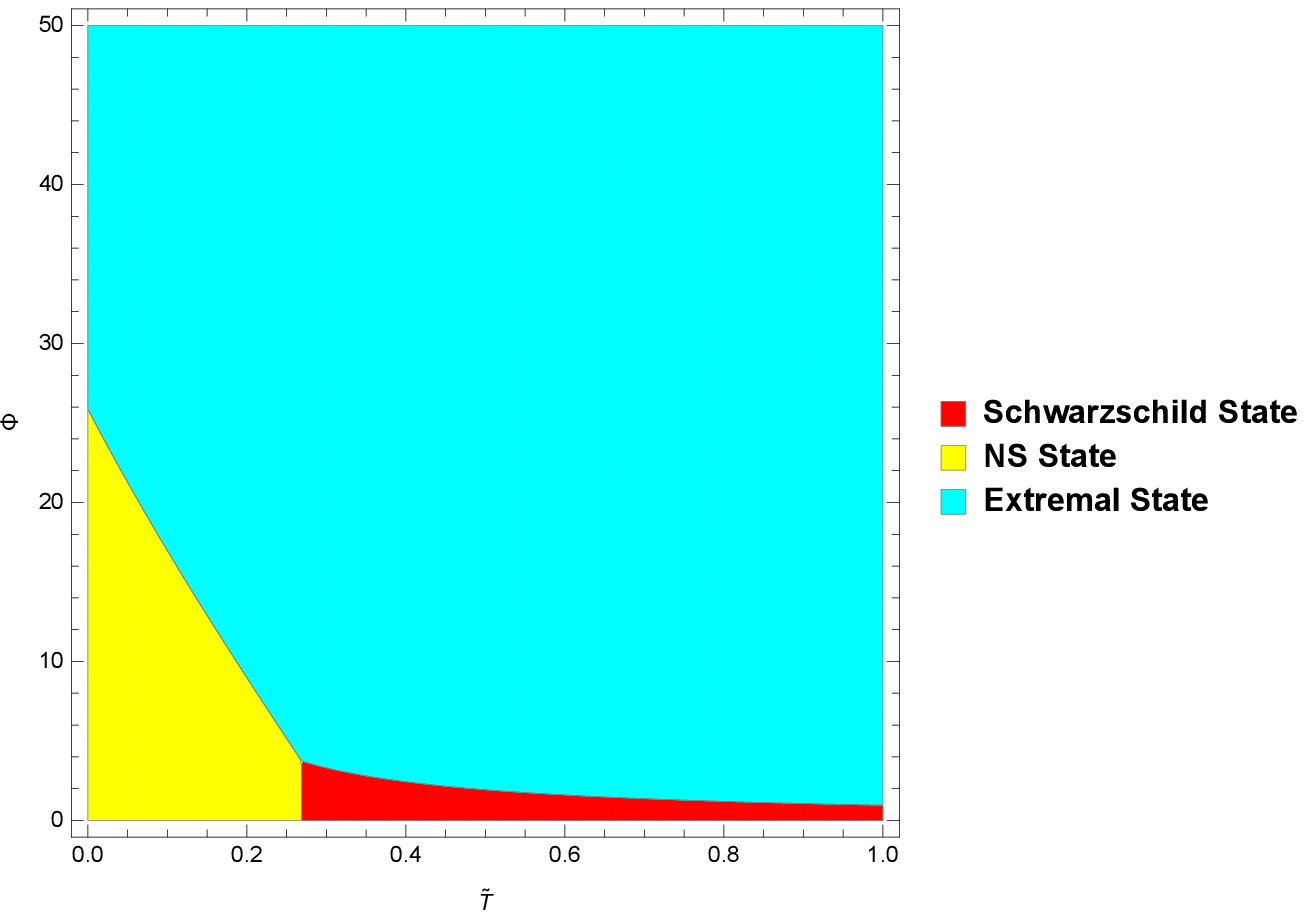}
\end{center}
\caption{{\footnotesize \textbf{Left Panel}: Physically allowed values of $x$
and $\tilde{Q}$ for BI black holes in a cavity. The blue lines are the
boundaries, on which the colored dots represent the candidates for the global
minimum state of the free energy on the boundaries. \textbf{Right Panel}:
Global minimum state of the free energy on the boundaries in the $\tilde{T}%
$-$\Phi$ space with $\tilde{a}=0.1$. Only NS State and Extremal State can be
the global minimum state on the whole physical region.}}%
\label{fig:CB}%
\end{figure}

For the BI black holes residing in a cavity, there appears to be some
constraints imposed on $x$ and $\tilde{Q}$. As shown in \cite{IN-Wang:2019kxp}%
, when $\tilde{Q}^{2}<4\tilde{a}$, BI black holes are Schwarzschild-like type,
which exist for $0<r_{+}<r_{B}$, or $0<x<1$ in tilde variables. When
$\tilde{Q}^{2}\geq4\tilde{a}$, BI black holes are RN type, which can have the
extremal BI black hole solution with the nonzero horizon radius $r_{e}%
=\sqrt{Q^{2}-4a}/2$. Requiring that $r_{e}<r_{+}<r_{B}$ leads to $\sqrt
{\tilde{Q}^{2}-4\tilde{a}}/2<x<1$ and $\tilde{Q}^{2}\leq4\left(  1+\tilde
{a}\right)  $. The physically allowed region for $x$ and $\tilde{Q}$ is
depicted as the gray area in the left panel of FIG. \ref{fig:CB}. To determine
the phase structures and transitions of a BI black hole residing in a cavity,
we should find the local and global minima of the free energy over the
physically allowed region of $x$ and $\tilde{Q}$. Solve eqn. $\left(
\ref{eq:LE}\right)  $ for $x$ and $\tilde{Q}$ gives the possible local minima
of the free energy in the $x$-$\tilde{Q}$ space. However, one also needs to
evaluate the free energy on the boundaries of the physically allowed region of
$x$ and $\tilde{Q}$ to determine the global minimum: \ 

\begin{itemize}
\item $B1$: $0\leq x\leq1$ and $\tilde{Q}=0$. The global minimum of the free
energy on $B1$ is at $x=0$ when $\tilde{T}<T_{B1c}\approx0.2686$ and at
$x=x_{\text{B1min}}>0$ otherwise. When $\tilde{Q}=0$ and $x=0$, the boundary
state is just the thermal flat space. The boundary state with $\tilde{Q}=0$
and $x=x_{\text{B1min}}$ is a Schwarzschild black hole in a cavity, which is
dubbed as Schwarzschild State. However, one finds that%
\begin{equation}
\frac{\partial\tilde{F}}{\partial Q}|_{B1}=-\frac{\Phi}{16\pi}<0,
\end{equation}
which means that neither the thermal flat space nor Schwarzschild State can be
the global minimum of the free energy over the whole physical region of $x$
and $\tilde{Q}$.

\item $B2$: $x=0$ and $0\leq\tilde{Q}\leq2\sqrt{\tilde{a}}$. For the state on
$B2$, its metric and Ricci scalar are%
\begin{equation}
f\left(  r\right)  =1-\frac{Q}{2\sqrt{a}}+\mathcal{O}\left(  r\right)  \text{
and }R=\frac{Q}{\sqrt{a}r^{2}}-\frac{2}{a}+\mathcal{O}\left(  r\right)
\text{,}%
\end{equation}
respectively. Although the metric is regular, the spacetime has a physical
singularity at $r=0$. So the state on $B2$ is a naked singularity since it has
no horizon. The global minimum of the free energy on $B2$ is at $\tilde
{Q}=\tilde{Q}_{\text{B2min}}$ with $0<\tilde{Q}_{\text{B2min}}<2\sqrt
{\tilde{a}}$ when $\Phi\leq\Phi_{\text{B2c}}$ and at $\tilde{Q}=2\sqrt
{\tilde{a}}$ otherwise. For simplicity, we denote the boundary state at
$\tilde{Q}=\tilde{Q}_{\text{B2min}}$ and $x=0$ as NS State. Note that the
thermal flat space, which is at $\tilde{Q}=0$ and $x=0$ on $B2$, is never the
global minimum of the free energy on $B2$.

\item $B3$: $x=\frac{1}{2}\sqrt{\tilde{Q}^{2}-4\tilde{a}}$ and $2\sqrt
{\tilde{a}}\leq\tilde{Q}\leq2\sqrt{1+\tilde{a}}$. The boundary state on $B3$
is an extremal BI black hole. In particular, the boundary state at $\tilde
{Q}=2\sqrt{1+\tilde{a}}$ and $x=1$ corresponds to the extremal BI black hole
with the horizon merging with the wall of the cavity. We denote this state as
Extremal State. It shows that the free energy can not have a local minimum on
$B3$ so the global minimum of the free energy on $B3$ is either at $\tilde
{Q}=2\sqrt{\tilde{a}}$ and $x=0$ or $\tilde{Q}=2\sqrt{1+\tilde{a}}$ and $x=1$.

\item $B4$: $x=1$ and $0\leq\tilde{Q}\leq2\sqrt{1+\tilde{a}}$. For a black
hole on $B4$, the event horizon merges with the wall of the cavity. On $B4$,
$\partial\tilde{F}/\partial Q=-\Phi/16\pi<0$, and hence the global minimum of
the free energy on $B4$ is at $\tilde{Q}=2\sqrt{1+\tilde{a}}$, which
corresponds to Extremal State.
\end{itemize}

We find that the global minimum of the free energy on the four boundaries can
only occur at Schwarzschild State, NS State or Extremal State, depending on
the values of $\tilde{a}$, $\tilde{T}$ and $\Phi$. In the right panel of FIG.
\ref{fig:CB}, the global minimum state on the boundaries is plotted in the
$\tilde{T}$-$\Phi$ space with $\tilde{a}=0.1$. As discussed above, only NS
State and Extremal State are the candidates for the global minimum state on
the physical region of $x$ and $\tilde{Q}$.

\begin{figure}[tb]
\begin{center}
\includegraphics[width=0.53\textwidth]{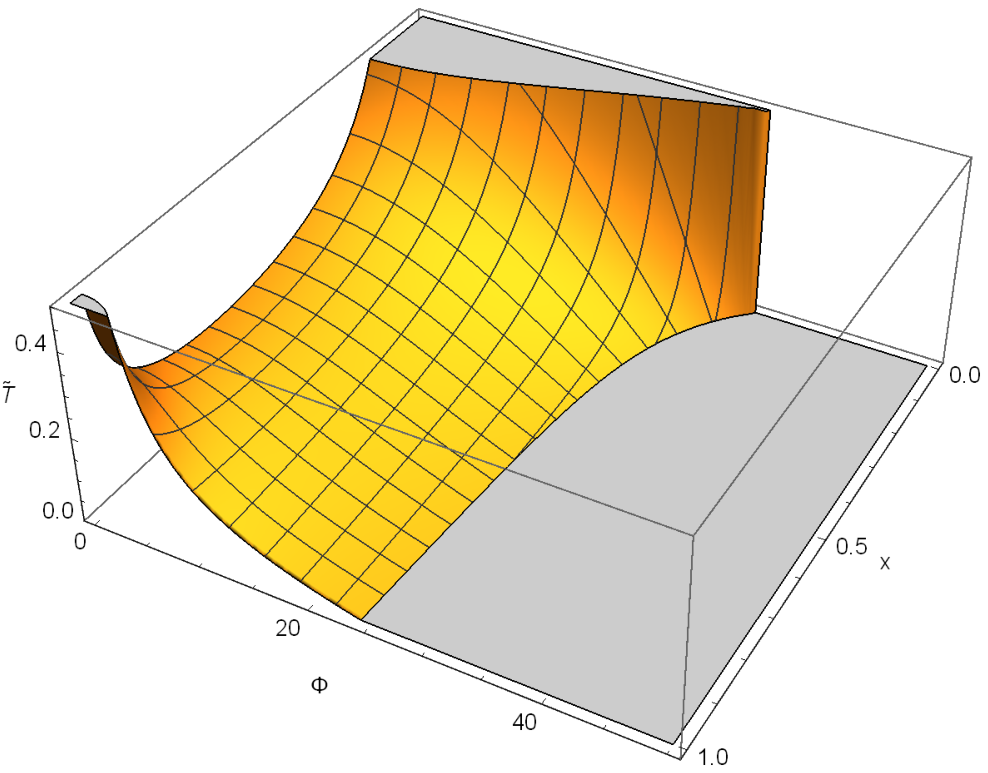}
\includegraphics[width=0.45\textwidth]{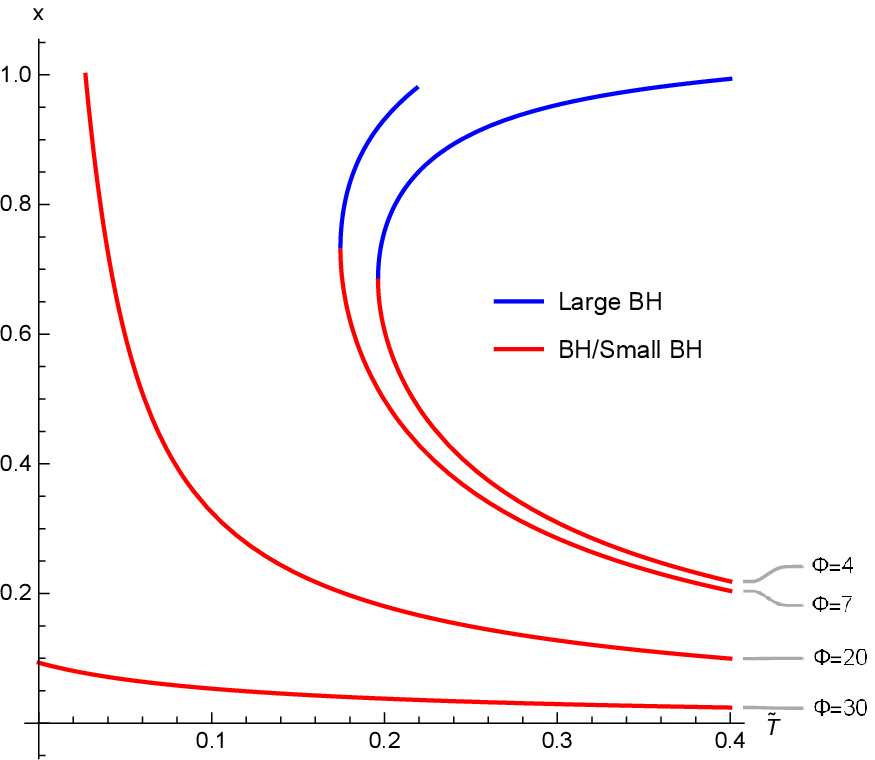}
\end{center}
\caption{{\footnotesize Plots of $\tilde{T}(x,\Phi)$ and $x(\tilde{T},\Phi)$
for BI black holes at locally stationary points. Here $\tilde{a}=0.1$.
\textbf{Left Panel}: Plot of $\tilde{T}$ as a function of $x$ and $\Phi$.
$\tilde{T}$ is negative in the gray area. \textbf{Right Panel}: Plot of $x$
against $\tilde{T}$ for various values of $\Phi$. The blue/red branches are
thermally stable/unstable.}}%
\label{fig:CT}%
\end{figure}

The black hole at the locally stationary points of the free energy can remain
in thermal equilibrium at constant temperature and potential in a cavity. To
determine the horizon radius of the black hole, we need to solve eqn. $\left(
\ref{eq:LE}\right)  $ for $x$ in terms of $\tilde{T}$ and $\Phi$: $x(\tilde
{T},\Phi)$. If $x(\tilde{T},\Phi)$ is multivalued, there are more than one
branch of different sizes. As $x\rightarrow0$, we find that there is a
critical potential $\Phi_{c1}$ such that $\tilde{T}\rightarrow+\infty\left(
-\infty\right)  $ when $\Phi<\Phi_{c1}\left(  \Phi>\Phi_{c1}\right)  $. As
$x\rightarrow1$, one has that
\begin{equation}
\tilde{T}\sim\frac{\Phi_{c2}^{2}-\Phi^{2}}{32\pi^{2}\sqrt{1+a}\Phi},
\end{equation}
where $\Phi_{c2}=\frac{8\sqrt{1+a}\pi}{\sqrt{1+2a}}<\Phi_{c1}$. Therefore at
$x=1$, $\tilde{T}>0\left(  <0\right)  $ when $\Phi<\Phi_{c2}\left(  \Phi
>\Phi_{c2}\right)  $. We also find that, for $\Phi>\Phi_{c1}$, $\tilde{T}$ is
always negative, and hence $\tilde{F}$ has no locally stationary points. When
$\Phi_{c2}<\Phi<\Phi_{c1}$, it can show that $\tilde{T}$ is monotonic as a
function of $x$, which means that there is only one branch of black holes with
fixed values of $\tilde{T}$ and $\Phi$. In the left panel of FIG.
\ref{fig:CT}, we plot $\tilde{T}$ as a function of $x$ and $\Phi$ with
$\tilde{a}=0.1$, which shows that $\tilde{T}$ is negative for large enough
value of $\Phi$, as expected. We also plot $x$ against $\tilde{T}$ for various
values of $\Phi$ with $\tilde{a}=0.1$ in the right panel of FIG. \ref{fig:CT}.
When $\Phi=4$ and $7$, there are two family of black holes of different sizes
with the same values of $\tilde{T}$ and $\Phi$: Small BH and and Large BH.
When $\Phi=20<\Phi_{c2}$, $\tilde{T}>0$ at $x=1$ and there is only one branch:
BH. When $\Phi=30>\Phi_{c2}$, there is still only one branch, on which there
exists an extremal black hole at $\tilde{T}=0$. Since thermally stable phases
have $\partial x/\partial\tilde{T}>0$, the BI black holes on blue/red branches
are thermally stable/unstable.

\begin{figure}[tb]
\begin{center}
\includegraphics[width=0.5\textwidth]{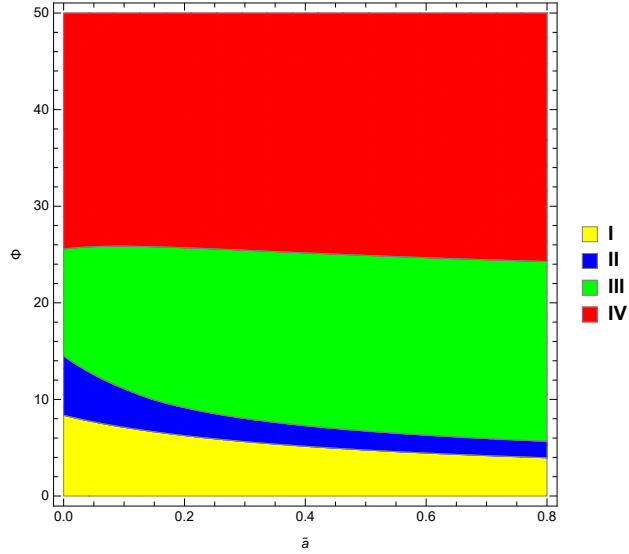}
\end{center}
\caption{{\footnotesize The four regions in the $\tilde{a}$-$\Phi$ space of
the systems with BI black holes enclosed in a cavity, each of which possesses
distinct behavior of the phase structures and transitions. Varying the
temperature, a first order NS State/Large BH phase transition and a second
order Large BH/Extremal State phase transition occur in Regions I, while only
a first order NS State/Extremal State phase transition occurs in Regions II
and III. There is no phase transition in Regions IV. }}%
\label{fig:CRP}%
\end{figure}\begin{figure}[ptb]
\begin{center}
\subfigure[{~\scriptsize Region I: $\tilde{a}=0.1$ and $\Phi=4$. As $\tilde{T}$ increases from zero, a first order phase transition
from NS State to Large BH occurs at the blue dot. Further increasing
$\tilde{T}$, there will be a second order phase transition from Large BH to
Extremal State at the brown dot.}]{
\includegraphics[width=0.45\textwidth]{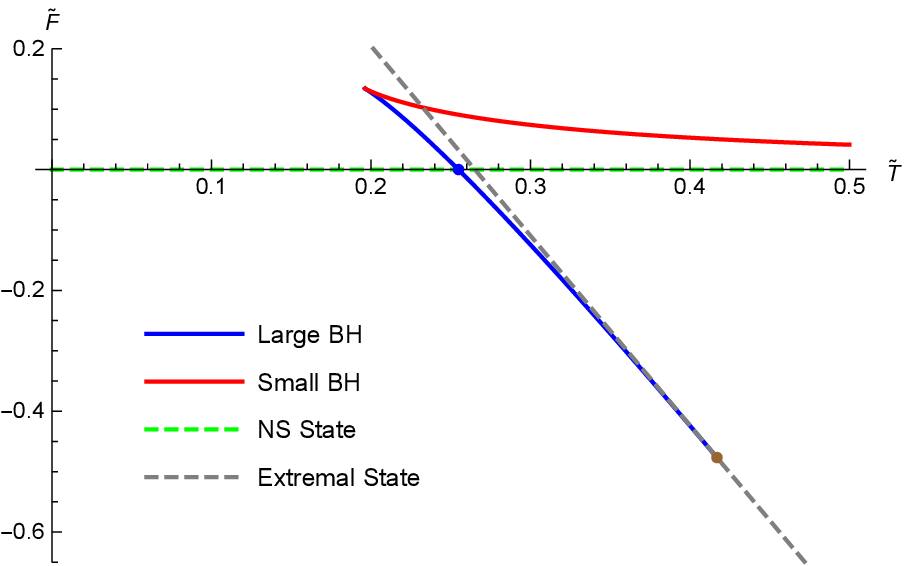}\label{fig:CR:a}}
\subfigure[{~ \scriptsize Region II: $\tilde{a}=0.1$ and $\Phi=7$. As $\tilde{T}$ increases from
zero, there is a first order phase transition at the blue dot from NS State to Extremal State. The second phase transition from Large BH to Extremal State is cloaked by NS State.}]{
\includegraphics[width=0.45\textwidth]{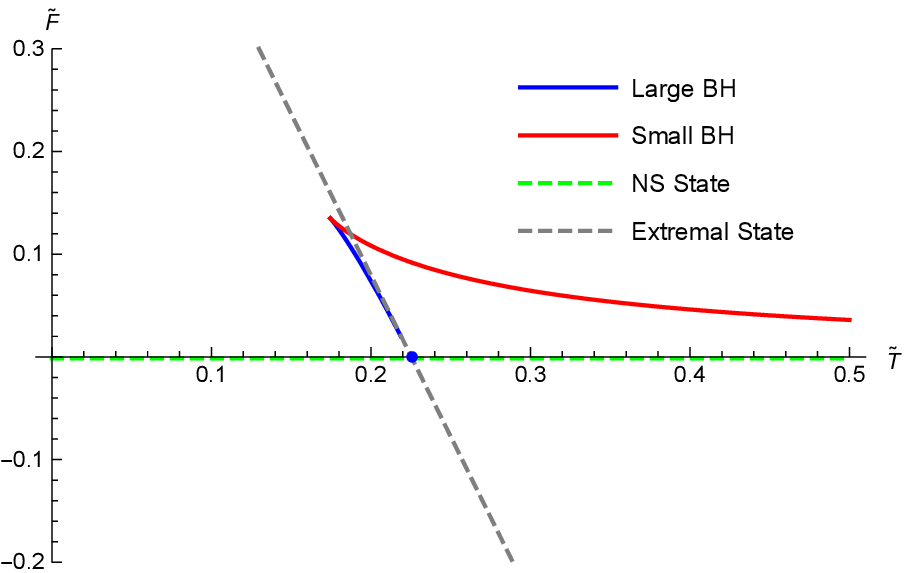}\label{fig:CR:b}}
\subfigure[{~ \scriptsize Region III: $\tilde{a}=0.1$ and $\Phi=20$. As the system is heated, it undergoes a first order phase transition from NS State to Extremal State.}]{
\includegraphics[width=0.45\textwidth]{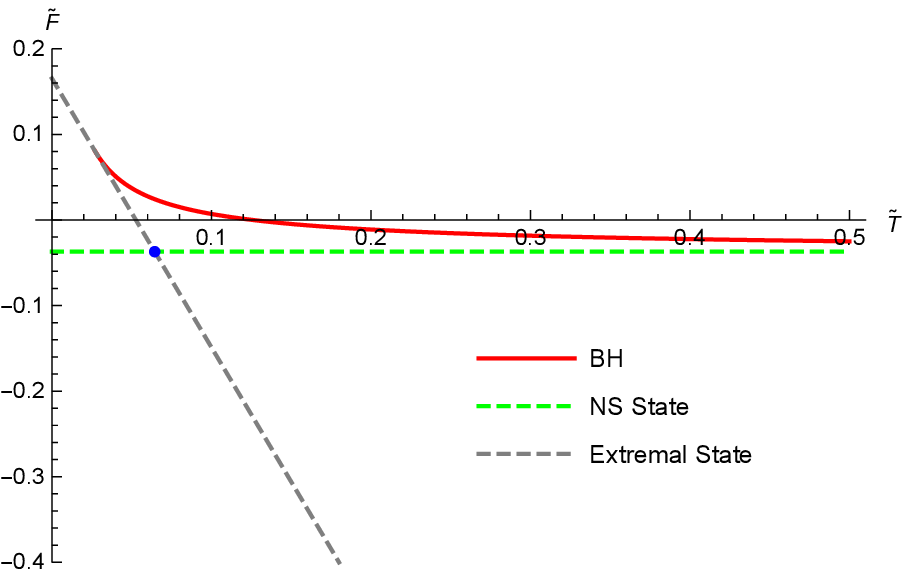}\label{fig:CR:c}}
\subfigure[{~ \scriptsize Region IV: $\tilde{a}=0.1$ and $\Phi=30$. The globally stable phase only is Extremal State, and hence there is no phase transition.}]{
\includegraphics[width=0.45\textwidth]{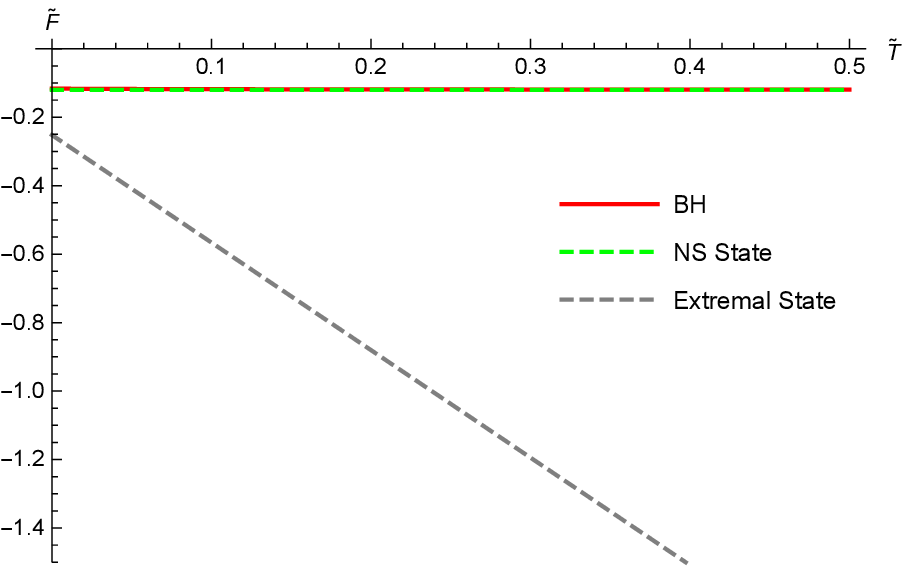}\label{fig:CR:d}}
\end{center}
\caption{{\footnotesize Plots of the free energy $\tilde{F}$ against the
temperature $\tilde{T}$ for different branches of BI black holes at the
locally stationary points, NS State and Extremal State of the systems in
Regions I, II, III and IV. The black holes on the blue branches are thermally
stable. The blue/brown dots represent first/second phase transitions of the
globally stable phases.}}%
\label{fig:CR}%
\end{figure}

\begin{figure}[tb]
\begin{center}
\includegraphics[width=0.5\textwidth]{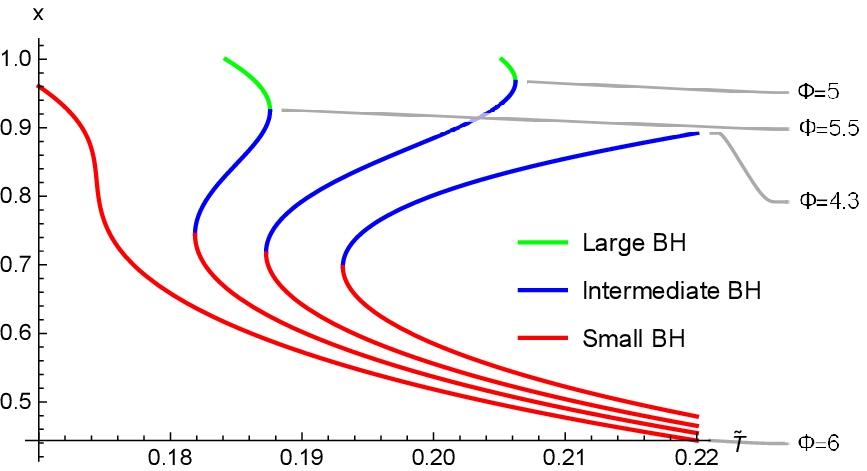}
\includegraphics[width=0.45\textwidth]{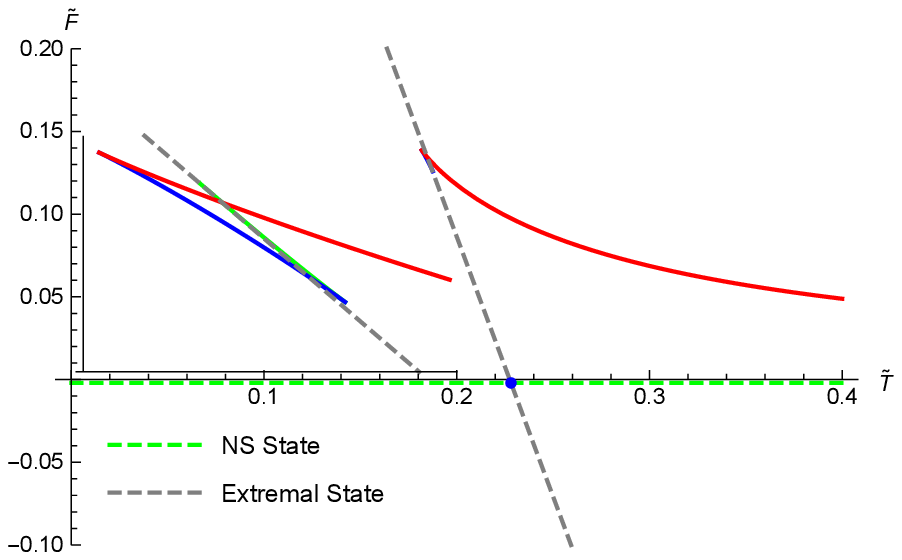}
\end{center}
\caption{{\footnotesize \textbf{Left Panel}: Plot of $x$ against $\tilde{T}$
for various values of $\Phi$ with $\tilde{a}=0.7$. Only the blue branches are
thermally stable. The systems with $\Phi=5$ and 5.5 are in Region II, which
have three branches of BI black holes. \textbf{Right Panel}: Plot of
$\tilde{F}$ against $\tilde{T}$ for the phases of the system with $\Phi=5.5$
and $\tilde{a}=0.7.$ As $\tilde{T}$ increases, the system undergoes a first
order transition from NS State to Extremal State. }}%
\label{fig:CR3}%
\end{figure}

Evaluating the free energy both on the boundaries and at the locally
stationary points, we find that there are four regions in the $\tilde{a}%
$-$\Phi$ phase space, in each of which the the system have different phase
structures and transitions. These four regions of the $\tilde{a}$-$\Phi$ phase
space are mapped in FIG. \ref{fig:CRP}. In what follows, we discuss the system
in the four regions:

\begin{itemize}
\item Region I: There is a temperature of $\tilde{T}_{\min}>0$, above which
black holes at the locally stationary points have two branches: Large BH and
Small BH. The Large/Small BH branch is thermally stable/unstable. The free
energies of the two branches, NS State and Extremal State are plotted in FIG.
\ref{fig:CR:a}, where $\tilde{a}=0.1$ and $\Phi=4$. For $\tilde{T}<\tilde
{T}_{\min}$, there are no locally stationary points, and the global minimum of
the free energy is at NS State. At $\tilde{T}=\tilde{T}_{\min}$, locally
stationary points start to appear. As $\tilde{T}$ increases from $\tilde
{T}_{\min}$, the free energy of Large BH decrease while that of NS state is
constant. They cross each other at the blue dot, where a first order phase
transition occurs. Further increasing $\tilde{T}$, Large BH stays globally
stable until it terminates and merge into Extremal State at the brown dot. At
the brown dot, Large BH and Extremal State both have $x=1$, and the entropy is
continuous across the transition. So the phase transition occurring at the
brown dot is a second order one, after which Extremal State becomes the
globally minimum state.

\item Region II: As in Region I, the free energy has no locally stationary
points, and the global minimum state is NS state when $\tilde{T}<\tilde
{T}_{\min}$. The locally stationary points of the free energy emerge when
$\tilde{T}>\tilde{T}_{\min}$, which consist of one thermally stable branch and
one or two thermally unstable branches. When there are two branches of black
holes, the free energies of the two branches, NS State and Extremal State are
plotted in FIG. \ref{fig:CR:b}, where $\tilde{a}=0.1$ and $\Phi=7$. It shows
that there seems to be a second order phase transition from Large BH to
Extremal Sate. However, the Large BH branch and the second order phase
transition are never the global minimum. So we only have a first order phase
transition from NS State to Extremal State occurring at the blue dot. In this
region, the black holes at the locally stationary points could also have three
branches, which are plotted in the left panel of FIG. \ref{fig:CR3}. Only
Intermediate BH is thermally stable. The free energies of the three branches,
NS State and Extremal State are plotted in the right panel of FIG.
\ref{fig:CR3}, where $\tilde{a}=0.7$ and $\Phi=5.5$. It shows that, as
$\tilde{T}$ increases, the global minimum also experiences a first order phase
transition from NS State to Extremal State. The inset in the right panel of
FIG. \ref{fig:CR3} illustrates that the phase transition from Intermediate BH
to Extremal State is first order due to the existence of Large BH.

\item Region III: The locally stationary points only consist of one thermally
unstable branch, which can never be the global minimum. The free energies of
the unstable branch, NS State and Extremal State are plotted in FIG.
\ref{fig:CR:c}, where $\tilde{a}=0.1$ and $\Phi=20$. The global minimum state
is NS state for low temperatures and Extremal State for high temperatures. A
first order phase transitions occurs at the blue dot.

\item Region IV: When $\Phi<\Phi_{c1}$, the locally stationary points exist
and have only one branch of thermally unstable black holes. The free energies
of the unstable branch, NS State and Extremal State are plotted in FIG.
\ref{fig:CR:c}, where $\tilde{a}=0.1$ and $\Phi=30$. Extremal State is the
only global stable phase, and there is no phase transition. When $\Phi
>\Phi_{c1}$, there are no locally stationary points for the free energy, and
Extremal State is the only globally stable phase as well.
\end{itemize}

\begin{figure}[tb]
\begin{center}
\includegraphics[width=0.45\textwidth]{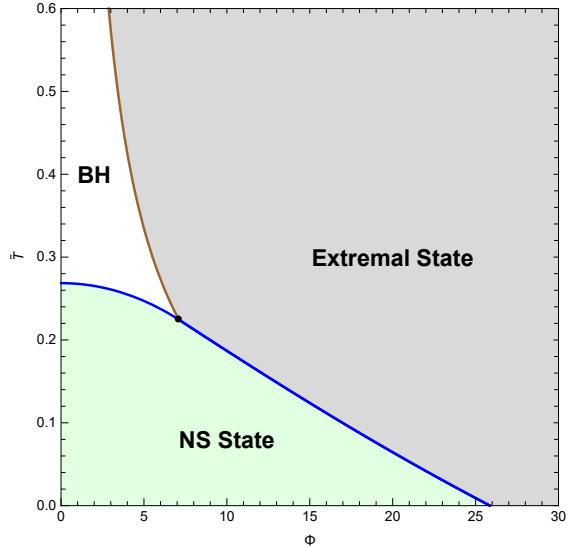}
\end{center}
\caption{{\footnotesize The globally stable phase diagram of the system with
$\tilde{a}=0.1$ in the $\Phi$-$\tilde{T}$ space. The blue/brown line
represents a first/second phase transition line. The three phase transitions
merge at the black dot. }}%
\label{fig:CPD}%
\end{figure}

In FIG. \ref{fig:CPD}, the globally stable phase diagram of the system with
$\tilde{a}=0.1$ is displayed in the $\Phi$-$\tilde{T}$ phase space. For NS
State and Extremal State, the system admits a globally stable phase on the
boundaries of the physical $x$ and $\tilde{Q}$ region. As discussed above,
there is no more than one thermally stable branch of BI black holes at the
locally stationary points. So the system admits at most one locally stable
phase, which describes a BI black hole in thermally stable equilibrium in a
cavity and is denoted by BH. The BH phase occurs in the phase diagram when it
is globally stable. There is a BH/NS State first order phase transition for
some range of $\Phi$, a NS State/Extremal State first order phase transition
for some smaller range of $\Phi$ and a BH/Extremal State second order phase
transition for some larger range of $\Phi$. These three phase transition lines
merge together at the black dot.

\section{Discussion and Conclusion}

\label{Sec:Con}

In this paper, we studied the phase structures and transitions of BI black
holes in a grand canonical ensemble by considering two boundary conditions,
namely the asymptotically AdS boundary and the Dirichlet boundary in the
asymptotically flat spacetime. For BI-AdS black holes, the phase structure
with respect to $\tilde{a}$ and $\Phi$ was displayed in FIG. \ref{fig:RP},
where there are four regions. For fixed values of the potential $\Phi$ and the
temperature $T$, the black holes in Regions I, II and III admit two solutions
of different sizes: Large BH (thermally stable) and Small BH (thermally
unstable). In Region IV, there is only one branch of black hole solutions,
which are thermally stable. In FIG. \ref{fig:PD}, the globally stable phases
and the phase transitions were shown in the $\Phi$-$\tilde{T}$ phase space.
There are two the globally stable phases, which are BH and thermal AdS space.
There are a BH/Thermal AdS zeroth order phase transition for some range of
$\Phi$ and a BH/Thermal AdS first order phase transition for smaller values of
$\Phi$. Note that the local and global stabilities of BI-AdS black holes in
the grand ensemble were already studied in \cite{IN-Fernando:2006gh}, where
the branches of BI-AdS black holes and the BH/Thermal AdS first and zeroth
order phase transitions were found. In this paper, we investigated phase
structures of BI-AdS black holes in the grand ensemble in a more thorough way.
To our knowledge, the phase diagrams \ref{fig:RP} and \ref{fig:PD} has yet to
be reported. Moreover, Region III of FIG. \ref{fig:RP} was not observed in
\cite{IN-Fernando:2006gh}. One can also study the phase structures and
transitions of BI-AdS black holes in the context of the extended phase space
thermodynamics, where the cosmological constant is interpreted as
thermodynamic pressure, i.e., $P=6/l^{2}$
\cite{Con-Dolan:2011xt,Con-Kubiznak:2012wp}. Our results can simply be
generalized to the extended phase space case by making replacements%
\begin{equation}
\tilde{T}=T\sqrt{6/P}\text{, }\tilde{a}=aP/6\text{ and }\tilde{F}\equiv
F\sqrt{P/6}.
\end{equation}

To determine the phase structure of BI black holes in a cavity, we computed
the locally stationary points of the free energy of the system over the
physical parameter space and the global minimum on the corresponding
boundaries. For the global minimum state on the boundaries, only NS State,
which describes a naked singularity, and Extremal State, which describes an
extremal black hole with the horizon merging with the wall of the cavity, are
the candidates for the global minimum state on the whole physical region. The
phase structure with respect to $\tilde{a}$ and $\Phi$ was displayed in FIG.
\ref{fig:CRP}, where there are also four regions. In Regions I and II, the
system admits one locally (thermally) stable phase while there are one or two
locally (thermally) unstable phases. The system in Region III and IV only has
one locally (thermally) unstable phase. The phases of the system that have the
globally minimum of the free energy were shown in FIG. \ref{fig:CPD}, which
are Black hole, NS State and Extremal State. The phase transitions between
globally stable phases of the system were also represented in FIG.
\ref{fig:CPD}, which shows there occur a Hawking-Page-like transition between
BH and NS State and a second-order phase transition between BH and Extremal
State. In this paper, we only focus on spherical topology, and hence it is
possible that there are some other states of lower free energy in a different
topological sector with the same potential and temperature. If this happens,
the globally stable phases discussed above could be only metastable.

For BI black holes in a cavity, the flat thermal space is on the boundary of
the physical region of the system. However, NS state or Extremal state is
always preferred over the flat thermal space. So the flat thermal space is
never the globally stable phase of the system. As shown in FIGs.
\ref{fig:CR:a} and \ref{fig:CR:b}, there are some regions of the parameter
space, in which NS state is globally stable while there is an unstable branch
of the BI black hole solution. Perturbing the unstable black hole, we find
that the black hole radiates away more energy than it absorbs, and the system
would eventually settle down to a naked singularity. Finding a time-dependent
solution, which describes a BI black hole evolving to a naked singularity, is
very tempting, since such solution can provide a counterexample to the weak
cosmic censorship conjecture \cite{Con-Penrose:1964wq}.

Finally, we found that, in the grand canonical ensemble, there are some
dissimilarities between the phase structures and transitions of BI-AdS black
hole and those of BI black holes in a cavity: 1) For BI-AdS black holes, the
thermal AdS space is sometimes preferred over the black hole solutions.
Inspired by the phase structure of RN black holes in a cavity
\cite{IN-Lundgren:2006kt}, one would expect that, for BI black holes in a
cavity, the thermal flat space could sometimes be globally preferred. However,
our results showed that the thermal flat space is never globally preferred.
Instead, NS state or Extremal State can be the globally minium state. 2)
Although a Hawking-Page-like first order phase transition occurs in both
cases, the system admits a second order phase transition for BI black holes in
a cavity and a zeroth order phase transition for BI-AdS black holes. 3) In
some regions in the parameter space of BI black holes in a cavity, the system
can have three locally extremal states of different sizes: one thermally
stable one and two thermally unstable ones. On the other hand, BI-AdS black
hole solutions have no more than two branches of different sizes.

\begin{acknowledgments}
We are grateful to Zheng Sun and Zhipeng Zhang for useful discussions and
valuable comments. This work is supported in part by NSFC (Grant No. 11005016,
11175039 and 11375121).
\end{acknowledgments}

\end{document}